Effects of equivalent composition on superconducting properties of high-entropy $RE$OBiS$_2$ ($RE$ = La, Ce, Pr, Nd, Sm, Gd) single crystals

Yuma Fujita[a], Masanori Nagao[a*], Akira Miura[b], Daisuke Urushihara[c], Yoshikazu Mizuguchi[d], Yuki Maruyama[a], Satoshi Watauchi[a], Yoshihiko Takano[e], and Isao Tanaka[a]

[a]*University of Yamanashi, 7-32 Miyamae, Kofu, Yamanashi 400-0021, Japan*

[b]*Hokkaido University, Kita-13 Nishi-8, Kita-Ku, Sapporo, Hokkaido 060-8628, Japan*

[c]*Nagoya Institute of Technology, Gokiso-cho, Showa-ku, Nagoya, Aichi 466-8555, Japan*

[d]*Department of physics, Tokyo Metropolitan University, 1-1, Minami-osawa, Hachioji, 192-0397, Japan*

[e]*National Institute for Materials Science, 1-2-1 Sengen, Tsukuba, Ibaraki 305-0047, Japan*

*Corresponding Author

Masanori Nagao

Postal address: University of Yamanashi, Center for Crystal Science and Technology

Miyamae 7-32, Kofu, Yamanashi 400-0021, Japan

Telephone number: (+81)55-220-8610

Fax number: (+81)55-220-8270

E-mail address: mnagao@yamanashi.ac.jp

**Abstract**

Superconductors are influenced by high-entropy alloys (HEAs); these have been investigated in various functional materials. $RE$OBiS$_2$ ($RE$ = La, Ce, Pr, Nd, Sm, and Gd in different combinations) single crystals with HEAs at the $RE$-site were successfully grown using the flux method. The obtained crystals were plate-shaped (~1 mm$^2$) with a well-developed $c$-plane. Ce was present in both trivalent (Ce$^{3+}$) and tetravalent (Ce$^{4+}$) electronic configurations; the concentration of Ce$^{4+}$ at the $RE$-site was approximately 10 at% in all single crystals. The single crystals showed superconducting transition temperature with zero resistivity within 1.2–4.2 K. The superconducting transition temperature, superconducting anisotropy, electronic specific heat coefficient, and Debye temperature of the crystals were not correlated with the mixed entropy at the $RE$-site. Except for the electronic specific heat coefficient, the variation of these parameters as a function of mixed entropy showed different trends for equivalent and non-equivalent $RE$ element compositions. Thus, the configuration of $RE$ elements influences the superconducting properties of $RE$OBiS$_2$ single crystals, alluding to a method of modulating transition temperatures.

# 1. Introduction

High-entropy alloys (HEAs) comprise five or more elements with concentrations between 5 and 35 at% [1]. HEAs have been extensively studied in various fields of functional materials [2,3]. Recently, enhanced superconducting properties by high-entropy strategy, such as the possible robustness of their superconducting states under extreme conditions (e.g., high pressure) [4], attracted great attention [5-7]. Although such materials with HEA concept introduced are extensively studied, fewer studies on the synthesis and characterization of single crystals [8-11]. Therefore, the anisotropic properties of the high-entropy effect have been limited.

$RE$OBiS$_2$ ($RE$: rare-earth elements) are BiS$_2$-based layered superconductors comprising alternating stacks of BiS$_2$ and $RE$O layers [12]. The superconductivity of BiS$_2$-based compounds can be induced by carrier doping and/or in-plane chemical pressure [13]. Carrier doping can be achieved by doping $O^{2-}$ sites with $F^-$ [14-18] or by performing valence fluctuation at the $RE$-site, for example, with Ce$^{3+}$ and Ce$^{4+}$ [19]. The superconducting transition temperatures of F-doped $RE$OBiS$_2$ are approximately 2−5 K and depend on both the fluorine concentration and substituted $RE$ elements [20].

In $RE$OBiS$_2$ superconductors without fluorine doping, the $RE$-site elements determine the superconducting transition temperature. The $RE$-site of $RE$OBiS$_2$ superconductors can be substituted for various *RE* elements, such as La+Ce, Ce+Pr, and Ce+Nd [21-23]. According to previous reports [5,6,8], *RE*OBiS$_2$ with the *RE*-site substituted by more than 5 kinds of rare earth elements have been referred to as *RE*OBiS$_2$ superconductors introducing the HEA concept. Recently, we grew single crystals of an HEA-type *RE*OBiS$_2$ (*RE* = La+Ce+Pr+Nd+Sm) using the flux method [8]. The superconducting properties of HEA-type *RE*OBiS$_2$ single crystals are affected by the chemical pressure effect [13] arising from the mean ionic radius of the *RE*-site. However, our previous report of HEA-type *RE*OBiS$_2$ single crystals did not involve a systematic investigation of the effect of the HEA concept on superconductivity.

Herein, a series of HEA-type *RE*OBiS$_2$ materials with nominal compositions that have the same mean *RE*-site ionic radius was employed to systematically investigate the effect of the HEA concept on the superconducting properties. We successfully grew HEA-type (La,Ce,Pr,Nd,Sm,Gd)OBiS$_2$ single crystals with different mixed entropies and similar mean *RE*-site ionic radii using the flux method. We also investigated the effect of the HEA concept on the superconducting transition temperature, superconducting anisotropy, and specific heat.

## 2. Material and methods

(La,Ce,Pr,Nd,Sm,Gd)OBiS$_2$ single crystals were grown using a high-temperature flux method [8, 24]. The starting materials used were La$_2$S$_3$, Ce$_2$S$_3$, Pr$_2$S$_3$, Nd$_2$S$_3$, Sm$_2$S$_3$, Gd$_2$S$_3$, Bi$_2$S$_3$, and Bi$_2$O$_3$. The raw materials were weighed to obtain a nominal composition, (La$_a$Ce$_b$Pr$_c$Nd$_d$Sm$_e$Gd$_f$)OBiS$_2$ ($a+b+c+d+e+f = 1.0$). A CsCl or CsCl/KCl mixture was used as the flux for crystal growth. A mixture of raw materials (0.8 g) and the alkali metal chloride flux (5.0 g) was ground using a mortar and pestle and then sealed in an evacuated (~10 Pa) quartz tube. The quartz tube was heated at $T_{max}$ for 10 h, followed by cooling to $T_{end}$ at a rate of 1 °C/h. The sample names are listed in Table 1. CsCl was employed as the flux, except for samples #*RE*-2 and #*RE*-4-e, with $T_{max}$ and $T_{end}$ values of 950 and 650 °C, respectively. The CsCl/KCl mixture (molar ratio of CsCl:KCl = 5:3, the eutectic temperature is 616 °C) was employed as the flux for samples #*RE*-2 and #*RE*-4-e [25] with $T_{max}$ and $T_{end}$ values of 850 and 600 °C, respectively. Next, the quartz tube was spontaneously cooled to room temperature (10–30 °C). The cooled quartz tube was opened in air, and the obtained materials were washed and filtered with distilled water to remove the alkali metal chloride flux.

Scanning electron microscopy (SEM) was conducted using a TM3030 system from Hitachi High Technologies, Japan. The compositional ratio of the *RE*OBiS$_2$ single crystals was evaluated using energy-dispersive X-ray spectrometry (EDS) (Quantax 70, Bruker Corp., Germany). The analytical composition of each element was defined as $C_{RE}$ (*RE*: La, Ce, Pr, Nd, Sm, Gd). The obtained values were normalized using the atomic content obeying $\Sigma(C_{RE}) = 1.00$ to clarify the relationship between the nominal and analytical compositions. Subsequently, Bi and S compositions ($C_{Bi}$ and $C_S$, respectively) were estimated to the precision of two decimal places. X-ray diffraction (XRD, MultiFlex, Rigaku, Japan) with a $2\theta/\theta$ scan using Cu-K$\alpha$ radiation was employed to confirm the crystal structure and determine the *c*-axis lattice constant.

The valence state of Ce in the *RE*OBiS$_2$ single crystals was estimated using X-ray absorption fine structure (XAFS) spectroscopy at the Aichi XAS beamline with a synchrotron X-ray radiation source (BL5S1). For the XAFS spectroscopy samples, the single crystals were ground, mixed with boron nitride (BN) powder, and pressed into a pellet with a diameter of 4 mm.

Resistivity−temperature ($\rho-T$) characteristics were measured using the standard four-probe method in constant-current density (*J*) mode using a physical property measurement system (PPMS DynaCool, Quantum Design, USA). The electrical terminals were fabricated using Ag paste (4922N, Du Pont, USA). The resistivity criterion of the superconducting transition temperature with zero resistivity ($T_c^{zero}$) was defined as 60 μΩ·cm. The superconducting anisotropy ($\Gamma_s$) was estimated by measuring the angular ($\theta$) dependence of resistivity ($\rho$) in the liquid-state flux under various magnetic fields ($H$) and applying an effective mass model [26].

The magnetization-temperature ($M$-$T$) characteristics under zero-field cooling (ZFC) and field cooling (FC) of the grown single crystals were measured using a superconducting quantum interface device (SQUID) magnetometer (MPMS, Quantum Design, USA), which featured the applied field of 10 Oe parallel to the $c$-axis. Magnetization was converted to magnetic susceptibility ($4\pi\chi$) using the density of CeOBiS$_2$ single crystal in the previous report [27].

The temperature dependence of the specific heat in the absence of a magnetic field was measured in the temperature range of 1.8−15 K by a mitigation method using a PPMS. The weight of the $RE$OBiS$_2$ single crystals for specific-heat measurements was in the range of 2.23−8.23 mg. The $RE$OBiS$_2$ single crystals were fixed on the sample stage using vacuum grease (Apiezon N grease, M&I Materials, UK). The addendum heat capacity was measured in a separate run and subtracted to obtain the sample heat capacity. The heating and cooling processes were performed thrice at the same temperature and the average values were recorded.

## 3. Results and discussion

*RE*OBiS$_2$ single crystals were grown from the typical flux method using slow cooling, and the heat treatment conditions of those single crystals were almost the same. Figure 1 shows a typical SEM image of the *RE*OBiS$_2$ single crystal with the HEA concept at the *RE*-site (sample #*RE*-6(Gd)). The single crystals have plate-like shapes with dimensions of 1 × 1 mm$^2$ and a thickness of approximately 100–300 μm. XRD measurements with the 2$\theta$/$\theta$ scan were performed on a well-developed plane of the single crystals. The presence of only 00*l* diffraction peaks, similar to that in tetragonal *RE*OBiS$_2$ [19], indicated the presence of a well-developed *c*-plane.

Table 1 shows that the nominal and analytical compositions of the *RE*OBiS$_2$ single crystals, and sample names. The analyzed compositions almost corresponded to the nominal compositions. *RE*OBiS$_2$ single crystals with more than 15 at% Gd substituted into the *RE*-site could not be grown under those growth conditions. It suggested that the *RE*OBiS$_2$ structure becomes unstable from Gd substitution which is the small ionic radius. The mixed entropy ($\Delta S_{mix}$) at the *RE*-site was calculated from the analytical compositions using the following equation [1]:

$$\Delta S_{mix} = -R \Sigma (C_{RE} \ln C_{RE}) \quad (RE: \text{La, Ce, Pr, Nd, Sm, Gd}) \quad \text{(Eq. 1)}$$

where $R$ = 8.314 J/mol·K is the gas constant. The concentration of each $RE$ element at the $RE$-site in samples #$RE$-5, #$RE$-5(Gd), #$RE$-5-e, and #$RE$-6(Gd) was within 8−32 at%, thereby meeting the definition of a HEA [1]. However, Cs, K, and Cl from the flux were not detected in the grown single crystals when the minimum sensitivity limit was approximately 1 wt%.

As shown in Table 2, the $c$-axis lattice constants of the $RE$OBiS$_2$ single crystals are estimated to be within 13.60−13.63 Å, except for that of CeOBiS$_2$ (#$RE$-1, 13.57 Å), by the XRD measurement with $2\theta/\theta$ scan. Therefore, the composition of the $RE$-site did not significantly influence the $c$-axis lattice constant. For $RE$OBiS$_2$ single crystals with the HEA concept at the $RE$-site, only Ce had a mixed valence state configuration, comprising trivalent (Ce$^{3+}$) and tetravalent (Ce$^{4+}$) electronic states [8]. The ratio of the Ce valence states (Ce$^{3+}$ and Ce$^{4+}$) at the $RE$-site was estimated using XAFS spectroscopy at room temperature. The Ce$^{4+}$ concentration at the $RE$-site was approximately 10 at% for all samples (Table 2). The mean $RE$-site ionic radius [29] was estimated by considering the valence state ($R^{3+}$ and $R^{4+}$) of the samples. As shown in Table 2, the mean $RE$-site ionic radii of all samples are within 1.101–1.109 Å, except for CeOBiS$_2$ (#$RE$-1, 1.127 Å). Surprisingly, the single crystals with equivalent $RE$ element compositions (i.e., #$RE$-2-e, #$RE$-3-e, #$RE$-4-e, and #$RE$-5-e) had the same range of *c*-axis lattice constants and mean *RE*-site ionic radius. Hence, it can be suggested that the stable *c*-axis lattice constants and mean *RE*-site ionic radius for the *RE*OBiS$_2$ crystal structure are within 13.60−13.63 and 1.101−1.109 Å, respectively. These results indicate that, except for #*RE*-1, the samples can be used to elucidate the effect of the HEA concept alone, without the influence of *RE*-site ionic radius and Ce$^{4+}$ concentrations effects.

Figure 2 shows the mixed entropy ($\Delta S_{\text{mix}}$) dependence of $T_c^{\text{zero}}$. The large increase in $T_c^{\text{zero}}$ from samples #*RE*-1 to #*RE*-2 is ascribed to the chemical pressure effect at the *RE*-site [13]. Samples #*RE*-5(Gd) and #*RE*-6(Gd) in the HEA region have lower $T_c^{\text{zero}}$ than those of the other samples. Note that both these samples are substituted by Gd at the *RE*-site. Therefore, the suppression of $T_c^{\text{zero}}$ in #*RE*-5(Gd) and #*RE*-6(Gd) can be explained by Gd substitution, as its presence may decrease the superconducting transition temperature in *RE*OBiS$_2$. A clear correlation between $\Delta S_{\text{mix}}$ and $T_c^{\text{zero}}$ is not observed. However, except for #*RE*-2-e, the single crystals with an equivalent *RE* element composition (i.e., samples #*RE*-3-e, #*RE*-4-e, and #*RE*-5-e), had lower $T_c^{\text{zero}}$ values than those of samples with non-equivalent *RE* element composition. In addition, $T_c^{\text{zero}}$ exhibits different behavior for the equivalent- and non-equivalent-composition samples. Therefore, the equivalent- and non-equivalent-composition samples were plotted in blue and red, respectively, in Figure 2.

We have previously reported that the $\Gamma_s$ values of $RE$OBiS$_2$ estimated using the upper critical field ratio and effective mass model were similar [8]. This indicates that both approaches are suitable to determine the superconducting anisotropies of $RE$OBiS$_2$. Therefore, we employed an effective mass model to estimate $\Gamma_s$. The reduced field ($H_{red}$) was calculated using the following equation:

$$H_{red} = H(\sin^2\theta + \Gamma_s^{-2}\cos^2\theta)^{1/2} \qquad \text{(Eq. 2)}$$

where $\theta$ is the angle between the $c$-plane and the magnetic field $H$ [30, 31]. $\Gamma_s$ was estimated from the best-scaling of the $\rho-H_{red}$ relationship. Figure 3 (a.1) and (b.1) shows the $\theta$ dependence of $\rho$ at various magnetic fields ($H$ = 0.01–9.0 T) in the liquid-state flux for samples (a.1) #$RE$-5 and (b.1) #$RE$-6(Gd) which are typical HEA-type samples. The $\rho-\theta$ curve exhibits an almost two-fold symmetry. Figure 3 (a.2) and (b.2) shows the $\rho-H_{red}$ scaling obtained from the $\rho-\theta$ curves shown in Figure 3 (a.1) and (b.1), respectively using Eq. 2. The scaling was determined for $\Gamma_s$ = 28 and 26, as shown in Figure 3 (a.2) and (b.2), respectively. The $\rho-\theta$ curves and $\rho-H_{red}$ scaling were similarly obtained for all samples. Figure 4 shows the relationship between $\Delta S_{mix}$ and $\Gamma_s$. The $\Gamma_s$ values also exhibit different behavior for the samples with equivalent and non-equivalent $RE$ element compositions. The $\Gamma_s$ values of the equivalent-composition samples (#*RE*-2-e, #*RE*-3-e, #*RE*-4-e, and #*RE*-5-e) were lower than those of the non-equivalent-composition samples. A similar trend was observed for $T_c^{zero}$. However, the equivalent- and non-equivalent-composition samples show different $T_c^{zero}$ and $\Gamma_s$ behavior. $\Gamma_s$ decreased in the HEA region for both sample types. Additionally, Gd substitution at the *RE*-site suppressed $T_c^{zero}$ but did not affect $\Gamma_s$. These results show that compared to $\Delta S_{mix}$, the superconducting properties ($T_c^{zero}$ and $\Gamma_s$) are significantly affected by an equivalent or non-equivalent *RE* element composition. A possible reason for this result is the different atom configurations of *RE* elements at the *RE*-site (ordering of *RE* elements). Hence, we will compare the lattice specific heat of the equivalent- and non-equivalent-composition samples at later.

Figure 5 shows the $4\pi\chi$-$T$ characteristic of samples (a) #*RE*-5 and (b) #*RE*-6(Gd), which exhibited the superconducting transition by the resistivity measurements. The transition temperatures in Figure 5 were consistent with the superconducting transition temperature in Figure 2. #*RE*-5 exhibited the $4\pi\chi$-$T$ characteristic with the typical superconductivity. The superconducting volume fraction of #*RE*-5 was approximately 65 %, which exhibits bulk superconductivity. But the exact shapes of those samples were not obtained, so the demagnetization factors have been not considered. On the other hand, the $4\pi\chi$-$T$ characteristic of #*RE*-6(Gd) showed the transition at approximately 3 K. However, it exhibited no-diamagnetism and paramagnetic behavior. Those behaviors may be due to Gd substitution and/or filamentary superconductivity. That cause was unclear from the performed magnetization measurements. The specific heat measurements of those samples (#*RE*-5 and #*RE*-6(Gd)) were performed. Figure 6 shows the temperature (*T*) dependence of the specific heat (*C*) (*C*–*T* curve) of samples (a) #*RE*-5 and (b) #*RE*-6(Gd). The squared temperature ($T^2$) dependence of *C* divided by *T* is shown in the inset of Figure 6 using the following expression:

$$C/T = \gamma + \beta T^2 \quad \text{(Eq. 3)}$$

where $\gamma$ is the electronic specific heat coefficient and $\beta$ is the coefficient for lattice specific heat. The red line in the inset of Figure 6 is a fit of the expression obtained using Eq. 3. The *C*/*T* value drastically increases near (a) 2.8 K ($T^2 = 7.84$) and (b) 2.9 K ($T^2 = 8.41$), may indicate a specific-heat jump for superconductivity. That temperature of (b) #*RE*-6(Gd) corresponded to approximately $T_c^{zero}$ of #*RE*-6(Gd), however that of (a) #*RE*-5 was lower than $T_c^{zero}$ of #*RE*-5 (approximately 3.7 K). The reason for the low temperature of specific-heat jump for #*RE*-5 is unclear. But the $4\pi\chi$-*T* characteristic of #*RE*-5 has shown bulk superconductivity, and then that specific-heat jump may be originated from superconductivity. The *C*–*T* curves exhibited similar behavior in both samples (a) #*RE*-5 and (b) #*RE*-6(Gd). Then it could not be explained that the $4\pi\chi$-*T* characteristic for #RE-6(Gd) originated from the Gd substitution and/or filamentary superconductivity.

From the inset of Figure 6, the values of $\gamma$ and $\beta$ were evaluated. The $\gamma$ values for #*RE*-5 and #*RE*-6(Gd) were 103 and 50.3 mJ/mol·K$^2$, respectively. Their $\beta$ values were 0.935 and 0.742 mJ/mol·K$^4$, respectively. In a simple Debye model of the phonon contribution, the $\beta$ coefficient is related to Debye temperature $\theta_D$:

$$\theta_D = \left(\frac{12\pi^4}{5\beta}nR\right)^{1/3} \quad \text{(Eq. 4)}$$

where $n = 5$ is the number of atoms per formula unit. From Eq. 4, $\theta_D$ of #*RE*-5 and #*RE*-6(Gd) were estimated to be 218 K and 236 K, respectively, which is higher than those reported for LaO$_{0.5}$F$_{0.5}$BiSSe single crystals (194 K) [32]. And they were similar to *RE*(O,F)BiS$_2$ with a single *RE* element at the *RE*-site of polycrystalline samples (189–222 K) [33]. Figure 7 shows the $\Delta S_{mix}$ dependence of (a) $\gamma$ and (b) $\theta_D$ estimated from the specific-heat measurement. The $\Delta S_{mix}$ dependence of $\gamma$ showed no appreciable correlation. However, $\theta_D$ drastically increased in the HEA region. Moreover, the $\theta_D$ of the equivalent-composition samples was lower than that of the non-equivalent-composition samples. These results indicate that the equivalent- and non-equivalent-composition samples have different lattice specific-heat contributions, which may become prominent in the HEA region. This behavior of $\theta_D$ is similar to that of $T_c^{zero}$ and $\Gamma_s$. Therefore, it can be suggested that the equivalent- and non-equivalent-composition samples show different behaviors for superconducting properties ($T_c^{zero}$ and $\Gamma_s$) owing to the lattice specific-heat contribution. In other words, the superconducting properties ($T_c^{zero}$ and $\Gamma_s$) of $RE$OBiS$_2$ superconductors are expected to be strongly affected by the lattice vibration state and phonon density of states, which are affected by the atom configuration (ordering).

The $RE$ elements in the equivalent-composition samples exhibit ordering at the $RE$-site. In contrast, the $RE$ element configuration in the non-equivalent-composition samples exhibits disordering at the $RE$-site. For example, non-equivalent-composition Ta$_{34}$Nb$_{33}$Hf$_8$Zr$_{14}$Ti$_{11}$ HEAs show superconductivity [34], whereas equivalent-composition Ta-Nb-Hf-Zr-Ti HEAs [35] do not exhibit superconductivity. We postulate the following regarding $RE$ element distribution: in the equivalent-composition samples, each $RE$ element was homogeneously substituted at the $RE$-site, whereas each $RE$-site was inhomogeneously substituted in the non-equivalent-composition samples. In the non-equivalent-composition samples, specific $RE$ elements become localized and introduce a local and/or correlated disorder [36], thereby affecting the superconducting properties ($T_c^{zero}$ and $\Gamma_s$). This phenomenon indicates that disordering at the $RE$-site increases the superconducting transition temperature of $RE$OBiS$_2$. Therefore, further characterization, such as via X-ray fluorescence holography [37, 38], is necessary to clarify the presence of local disorder in equivalent- and non-equivalent-composition samples.

## 4. Conclusions

The correlation between the $RE$-site mixed entropy ($\Delta S_{mix}$) and superconducting properties of $RE$OBiS$_2$ compounds was investigated by growing a series of (La,Ce,Pr,Nd,Sm,Gd)OBiS$_2$ single crystals of varying $RE$ element composition. Because the samples exhibited a similar $c$-axis lattice constant (13.60–13.63 Å) and mean $RE$-site ionic radius (1.101–1.109 Å), changes in superconducting properties can be ascribed to the HEA concept. The single crystals exhibited a superconducting transition temperature ($T_c^{zero}$) within 1.2–4.2 K, and a $\Delta S_{mix}$ dependence was not observed for $T_c^{zero}$ nor superconducting anisotropy ($\varGamma_s$). However, the $T_c^{zero}$ and $\varGamma_s$ values of samples with equivalent $RE$ element composition were lower than those of non-equivalent-composition samples. A similar phenomenon was observed for the Debye temperature ($\theta_D$), which reflects the lattice-vibration state and phonon density of states. Therefore, the atom configuration at the $RE$-site strongly affects the superconducting properties ($T_c^{zero}$ and $\Gamma_s$) of $RE$OBiS$_2$ compounds. We propose that the equivalent or non-equivalent composition at the HEA-site plays an important role in tuning the superconducting properties while the high-entropy effect (various the $RE$-site mixed entropy) on superconducting properties is less dominant. Furthermore, this research emphasizes the need to investigate the atomic ordering at the HEA-site.

**Acknowledgments**

The authors thank Dr. T. Yamamoto (NIMS) for useful discussions. The XAFS spectroscopy experiments were conducted at the BL5S1 of Aichi Synchrotron Radiation Center, Aichi Science & Technology Foundation, Aichi, Japan (Experimental No. 201905108, No. 202005002, and No. 202104005). This work was partially supported by JSPS KAKENHI (Grant-in-Aid for Scientific Research (B) and (C): Grant Number 21H02022 and 19K05248, Grant-in-Aid for Challenging Exploratory Research: Grant Number 21K18834). We would like to thank Editage (www.editage.com) for English language editing.

Table 1. Nominal and analyzed compositions and $\Delta S_{mix}$ of the $RE$-site in $RE$OBiS$_2$ single crystals.

| Sample | Nominal composition / Analytical composition | | | | | | | | $\Delta S_{mix}$ (R: Gas constant) |
|---|---|---|---|---|---|---|---|---|---|
| | La: $a$ $C_{La}$ | Ce: $b$ $C_{Ce}$ | Pr: $c$ $C_{Pr}$ | Nd: $d$ $C_{Nd}$ | Sm: $e$ $C_{Sm}$ | Gd: $f$ $C_{Gd}$ | Bi $C_{Bi}$ | S $C_S$ | |
| **#$RE$-1 [27]** | | 1.00 / 1.0(1) | | | | | 1.00 / 1.01(5) | 2.00 / 2.00 | 0 |
| **#$RE$-2** | | 0.40 / 0.39(1) | | 0.60 / 0.61(1) | | | 1.00 / 0.98(1) | 2.00 / 2.00(5) | 0.6687$R$ |
| **#$RE$-2-e [23, 28]** | | 0.50 / 0.50(1) | | 0.50 / 0.50(1) | | | 1.00 / 0.98(1) | 2.00 / 2.00(5) | 0.6931$R$ |
| **#$RE$-3** | | 0.30 / 0.32(1) | 0.30 / 0.29(1) | 0.40 / 0.39(1) | | | 1.00 / 0.99(2) | 2.00 / 1.99(3) | 1.091$R$ |
| **#$RE$-3-e** | | 0.33 / 0.33(1) | 0.33 / 0.33(1) | 0.33 / 0.33(1) | | | 1.00 / 1.00(6) | 2.00 / 2.12(7) | 1.098$R$ |
| **#$RE$-4** | | 0.30 / 0.27(1) | 0.40 / 0.40(1) | 0.20 / 0.21(1) | 0.10 / 0.12(2) | | 1.00 / 1.08(3) | 2.00 / 2.16(3) | 1.302$R$ |
| **#$RE$-4-e** | 0.25 / 0.26(1) | 0.25 / 0.24(1) | | 0.25 / 0.25(1) | 0.25 / 0.25(1) | | 1.00 / 0.97(4) | 2.00 / 1.97(9) | 1.386$R$ |
| **#$RE$-5** | 0.10 / 0.09(1) | 0.30 / 0.30(2) | 0.30 / 0.32(1) | 0.10 / 0.09(1) | 0.20 / 0.20(1) | | 1.00 / 1.01(5) | 2.00 / 2.10(8) | 1.481$R$ |
| **#$RE$-5(Gd)** | 0.20 / 0.20(1) | 0.30 / 0.31(1) | 0.30 / 0.30(1) | | 0.10 / 0.09(1) | 0.10 / 0.10(1) | 1.00 / 0.88(1) | 2.00 / 2.00(3) | 1.493$R$ |
| **#$RE$-5-e [8]** | 0.20 / 0.23(1) | 0.20 / 0.21(1) | 0.20 / 0.19(2) | 0.20 / 0.19(1) | 0.20 / 0.17(1) | | 1.00 / 0.96(3) | 2.00 / 2.17(3) | 1.598$R$ |
| **#$RE$-6(Gd)** | 0.20 / 0.21(1) | 0.30 / 0.31(1) | 0.20 / 0.21(1) | 0.10 / 0.08(1) | 0.10 / 0.08(1) | 0.10 / 0.11(1) | 1.00 / 0.95(2) | 2.00 / 2.07(5) | 1.665$R$ |

Table 2. $c$-axis lattice constants, $Ce^{3+}$ and $Ce^{4+}$ contents at *RE*-site, and the mean *RE*-site ionic radius for grown *RE*OBiS$_2$ single crystals.

| Sample name | $c$-axis lattice constant (Å) | $Ce^{3+}$ and $Ce^{4+}$ contents at *RE*-site | | Mean *RE*-site ionic radius (Å) (Calculated from the ratio of $Ce^{3+}$ and $Ce^{4+}$) |
|---|---|---|---|---|
| | | $Ce^{3+}$ | $Ce^{4+}$ | |
| #*RE*-1 | 13.57 [27] | 0.91 | 0.090 | 1.127 |
| #*RE*-2 | 13.62 | 0.30 | 0.088 | 1.105 |
| #*RE*-2-e [22,23] | 13.60 | 0.37 | 0.13 | 1.104 |
| #*RE*-3 | 13.62 | 0.23 | 0.090 | 1.109 |
| #*RE*-3-e | 13.61 | 0.23 | 0.097 | 1.101 |
| #*RE*-4 | 13.63 | 0.18 | 0.093 | 1.109 |
| #*RE*-4-e | 13.63 | 0.13 | 0.11 | 1.104 |
| #*RE*-5 | 13.62 | 0.18 | 0.12 | 1.102 |
| #*RE*-5(Gd) | 13.61 | 0.18 | 0.13 | 1.104 |
| #*RE*-5-e [8] | 13.63 | 0.12 | 0.095 | 1.104 |
| #*RE*-6(Gd) | 13.62 | 0.19 | 0.12 | 1.105 |

**Figure captions**

Figure 1. Typical SEM image of a $RE$OBiS$_2$ single crystal with HEA concept at $RE$-site (sample #$RE$-6(Gd)).

Figure 2. Mixed entropy ($\Delta S_{mix}$) dependence of $T_c^{zero}$.

Figure 3. Angular dependence of resistivity in flux liquid-state under various magnetic fields ($H$ = 0.01–9.0 T) for samples (a.1) #$RE$-5 at 3.0 K and (b.1) #$RE$-6(Gd) at 2.5 K. (a.2) and (b.2) reduced magnetic field $H_{red}$ dependence of resistivity scaled using the equation $H_{red} = H(\sin^2 \theta + \Gamma_s^{-2} \cos^2 \theta)^{1/2}$ and the data given in (a.1) and (b.1), respectively.

Figure 4. Mixed entropy ($\Delta S_{mix}$) dependence of $\Gamma_s$.

Figure 5. Temperature ($T$) dependence of magnetic susceptibility ($4\pi\chi$) under zero-field cooling (ZFC) and field cooling (FC) with an applied field ($H$) of 10 Oe parallel to the $c$-axis for samples (a) #$RE$-5 and (b) #$RE$-6(Gd).

Figure 6. Temperature ($T$) dependence of the specific heat ($C$) of samples (a) #$RE$-5 and (b) #$RE$-6(Gd). The inset shows the squared temperature ($T^2$) dependence of $C$ divided by $T$. The red line is a fit of the expression $C/T = \gamma + \beta T^2$.

Figure 7. Mixed entropy ($\Delta S_{mix}$) dependence of (a) $\gamma$ and (b) $\theta_D$.

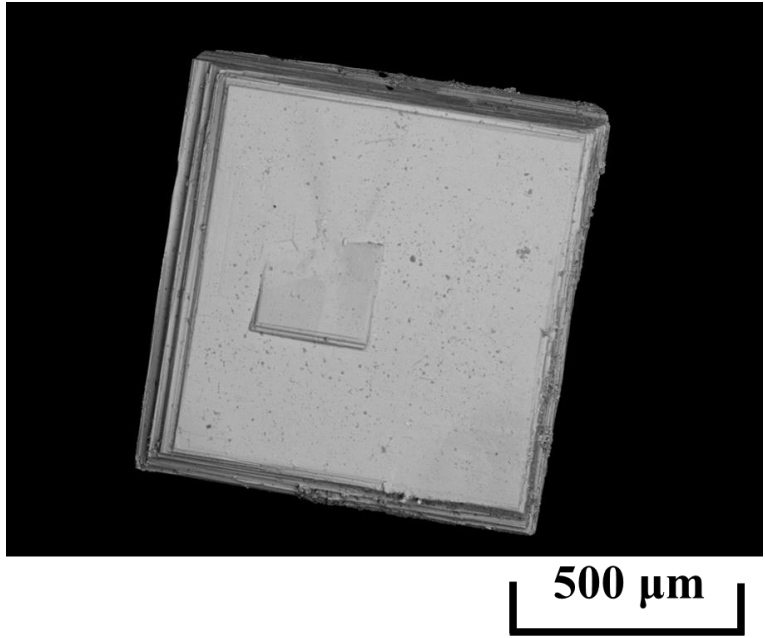

**Figure 1**

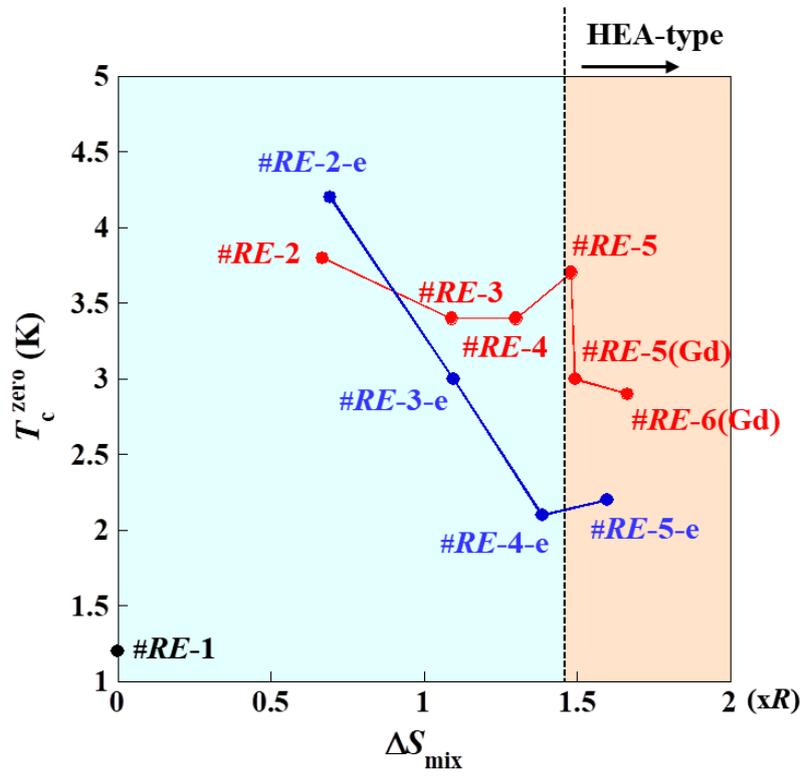

**Figure 2**

#*RE*-5

(a.1) 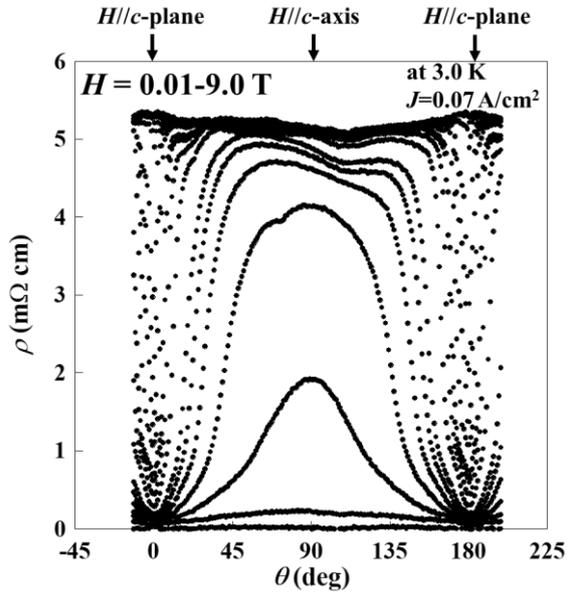

(a.2) 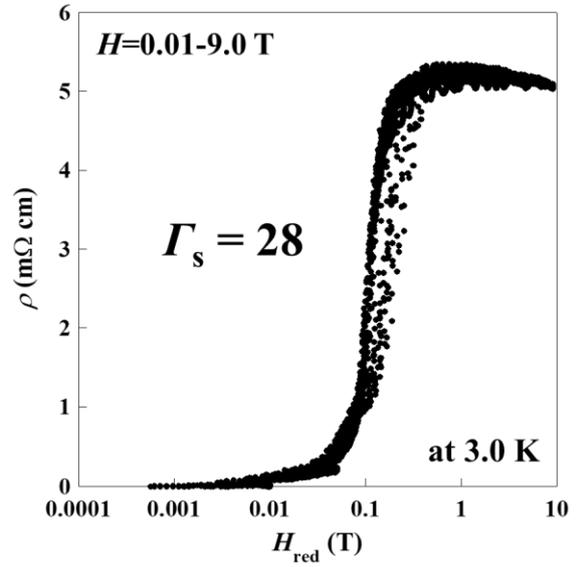

#*RE*-6(Gd)

(b.1) 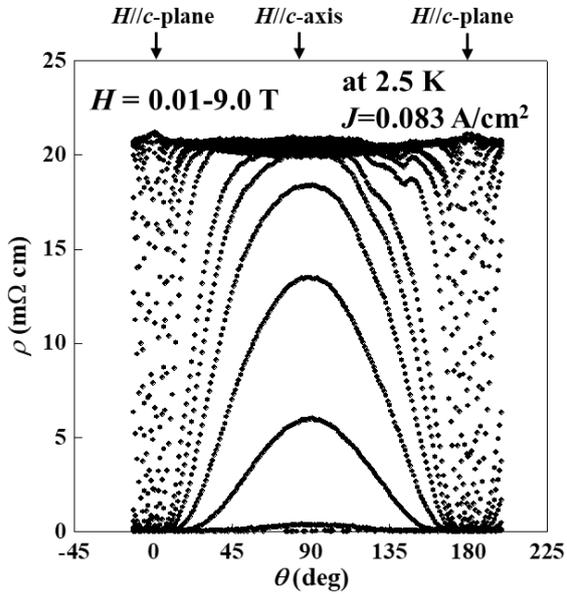

(b.2) 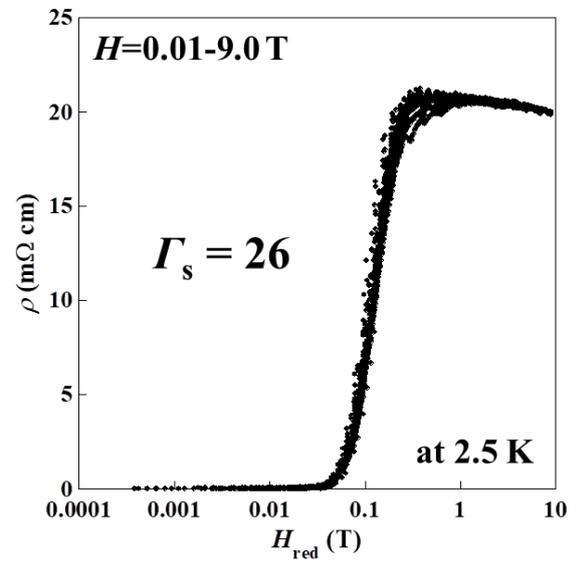

Figure 3

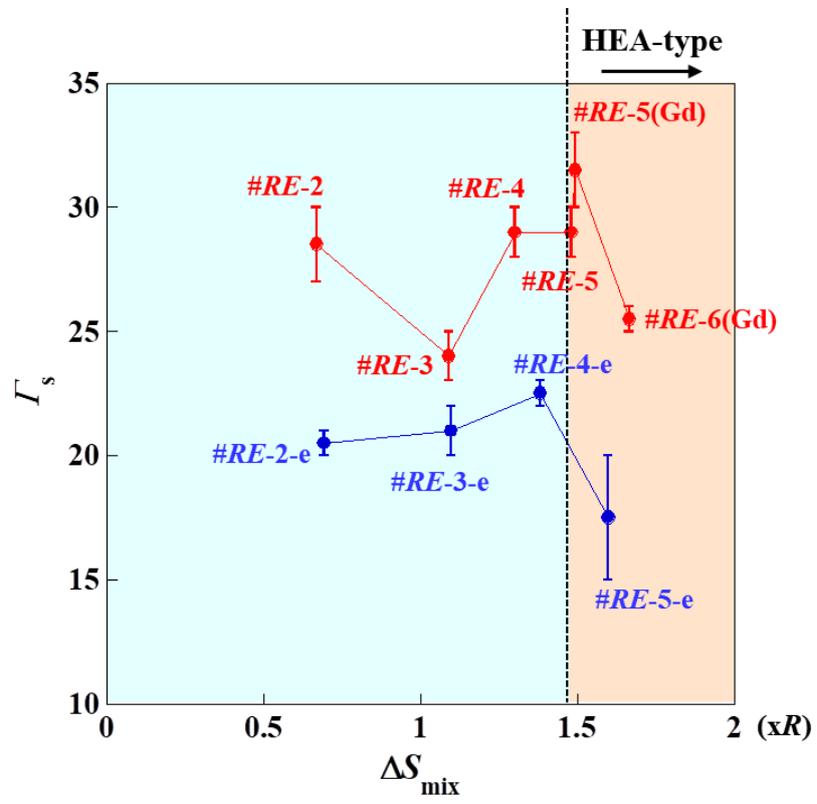

**Figure 4**

**(a) #*RE*-5**  **(b) #*RE*-6(Gd)**

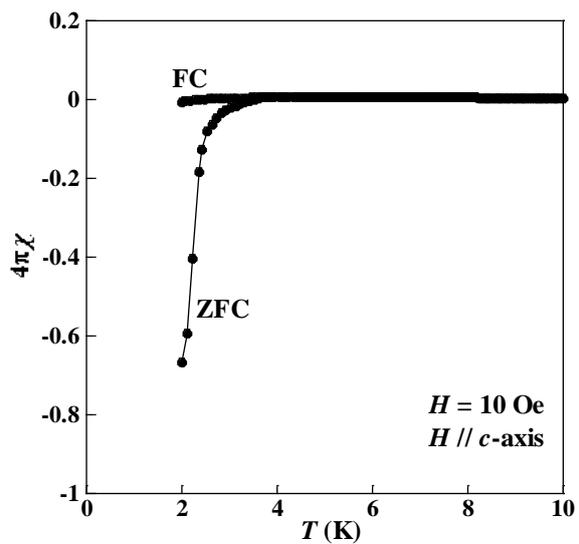 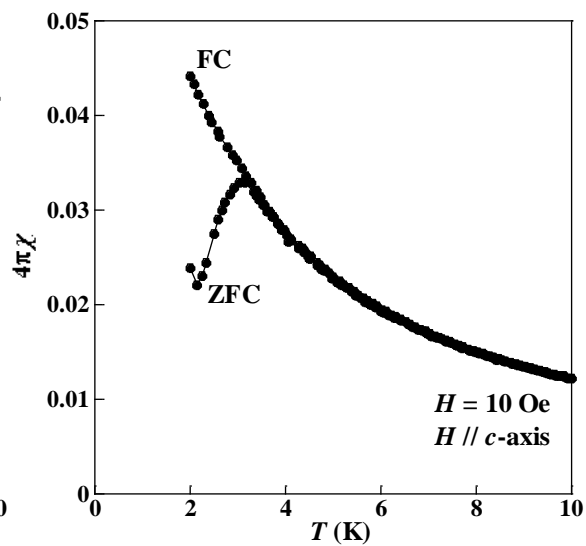

**Figure 5**

**(a)** *#RE*-5

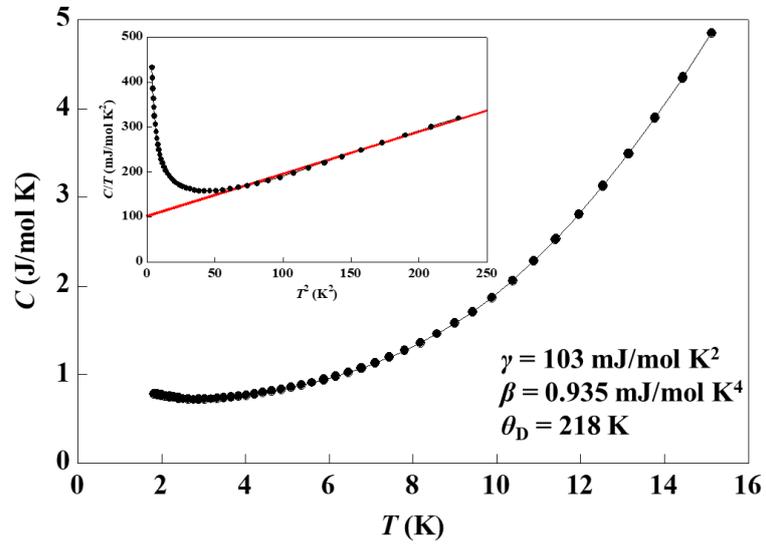

$\gamma$ = 103 mJ/mol K$^2$
$\beta$ = 0.935 mJ/mol K$^4$
$\theta_\mathrm{D}$ = 218 K

**(b)** *#RE*-6(Gd)

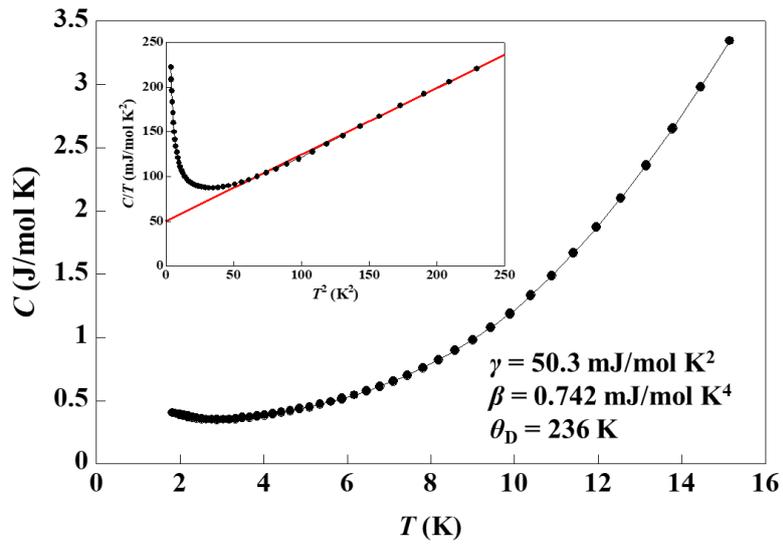

$\gamma$ = 50.3 mJ/mol K$^2$
$\beta$ = 0.742 mJ/mol K$^4$
$\theta_\mathrm{D}$ = 236 K

**Figure 6**

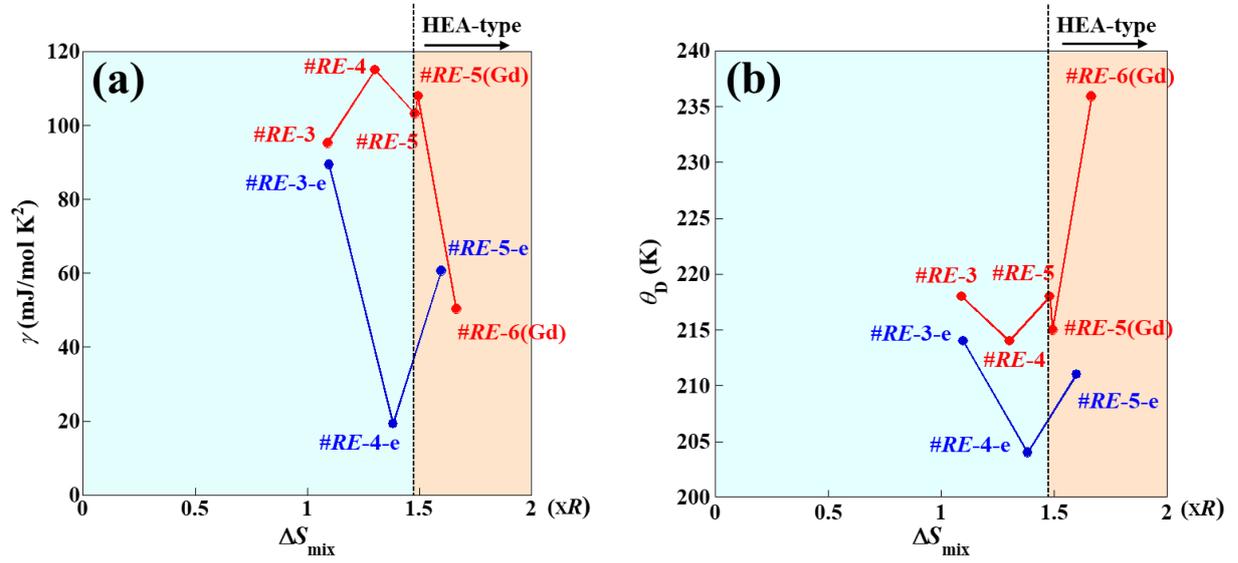

**Figure 7**